\documentclass[twocolumn,showpacs,prl,amsmath,amssymb]{revtex4}

\usepackage{graphicx}
\usepackage{amsmath}
\usepackage{amssymb}
\usepackage{bm}
\usepackage{bbm}
\usepackage{mathbbol}

\DeclareMathAlphabet{\mathpzc}{OT1}{pzc}{m}{it}
\DeclareMathOperator{\Real}{Re}

\newcommand{\bigO}[1]{\ensuremath{{\cal O}\left( #1 \right)}}

\newcommand{\clusterCountW}[1]{\hat{W}^{[#1]}}

\begin{document}

\title{Universal pulse sequence to minimize spin dephasing
in the central spin decoherence problem}
\author{B. \surname{Lee}$^{1,2}$, W. M. \surname{Witzel}$^{1,3}$,
  S. \surname{Das Sarma}$^{1}$}
\affiliation{$^{1}$Condensed Matter Theory Center,
  Department of Physics, University of Maryland, College Park, MD
  20742-4111, USA \\
$^{2}$Science, Mathematics, and Computer Science Magnet Program, Montgomery Blair High School, Silver Spring, MD
  20901, USA \\ 
$^{3}$Naval Research Laboratory, Washington, DC 20375, USA}
\date{\today}
\begin{abstract}
We present a remarkable finding that a recently discovered [G. S. Uhrig,
Phys. Rev. Lett. 98, 100504 (2007)] series
of pulse sequences, designed to optimally restore coherence to a
qubit in the spin-boson model of decoherence, is in fact completely
model-independent and generically valid for arbitrary dephasing
Hamiltonians given sufficiently short delay times between pulses.
The series maximizes qubit fidelity versus number of applied pulses
for sufficiently short delay times because the series, with each
additional pulse, cancels successive orders of a time expansion for
the fidelity decay. The ``magical'' universality of this property,
which was not appreciated earlier, requires that a linearly growing
set of ``unknowns'' (the delay times) must simultaneously satisfy an
exponentially growing set of nonlinear equations that involve
arbitrary dephasing Hamiltonian operators.
\end{abstract}

\maketitle


{\it Introduction.} The spin-boson (SB) model, where the environment is modelled by a
bosonic bath of simple harmonic oscillator quanta, is
a famous and extensively used general
technique for studying the quantum decoherence of a system coupled to
an environment~\cite{SpinBoson}.
The model is often used in the context of quantum dissipation and
decoherence analyses of wide classes of couplings between a system and
its environment.  
Recently, Uhrig~\cite{Uhrig} discovered a series of $\pi$-pulse sequences 
that optimally, with respect to the number of pulses,  
decouple a qubit from a bosonic bath in the SB model, thus protecting
the qubit from decoherence.
Because errors are inherently introduced by each pulse,
the promise of optimal dynamical decoupling (DD) is of
great significance to
the intense recent activity in quantum information processing where
minimizing qubit decoherence and successfully carrying out quantum
error correction protocols are crucial. 

A particularly important issue in view of the remarkable power of the
Uhrig DD (UDD) in fighting qubit decoherence is its applicability in
realistic situations beyond the idealized SB model context of its 
discovery~\cite{Uhrig}.
To understand the general applicability of these sequences, 
we set out to apply it to a drastically different model of
decoherence, the central spin
decoherence problem; specifically, we consider the spectral diffusion
(SD) process, quantum dephasing of an electron spin qubit coupled to a
slowly fluctuating nuclear spin environment (a spin bath).
Our investigations led us to discover that UDD transcends all models and is an
optimal decoupling sequence for any dephasing Hamiltonian
when delay times are sufficiently short because, with each additional
pulse, it kills successive orders of a time or Magnus~\cite{Magnus, MagnusFootnote}
expansion~\cite{footnote}.
The applicability of the UDD sequence to the SD model, with 
its extreme contrast to SB, already indicates its powerful generality. 
What we find is
much more general -- we show that the Uhrig sequence is model-independent.


How to preserve the state of a qubit in a bath is
an important theoretical and practical consideration
in the field of quantum information.  A large energy splitting between
the qubit's two logical states (e.g., through the
application of a magnetic field for a spin qubit) compared with the
temperature of the bath may result in long relaxation ($T_1$) times; however,
the relative phase of a superposition state may
not be preserved by this strategy so that dephasing ($T_2$) decoherence
ensues.
As a strategy to combat dephasing and an example of DD, a sequence of $\pi$-pulses may be
applied in order to rapidly, on the time scale of the system dynamics,
flip the qubit in between time intervals of free evolution (this may be generalized for any
quantum system using inverting~\cite{Dhar} pulses as a generalization of $\pi$-pulses).
In the simplest case, the Hahn spin echo occurs
after applying a single $\pi$ pulse midway through the 
system's evolution.  Concatenated~\cite{Khodjasteh} DD
(CDD) sequences can successively
improve coherence times, but at the considerable overhead 
expense of exponentially increasing the
number of applied pulses.  UDD performs its magic, not only in the SB
model but whenever delay times are sufficiently short,
with a mere linear scaling
in the number of applied pulses.


{\it Pulse sequence echoes.} 
Considering only dephasing decoherence, the effective Hamiltonian (for
any model) may be written in the form 
$\hat{\cal H} = \sum_{\pm} \lvert \pm \rangle \hat{\cal H}_{\pm} \langle \pm
\rvert$ with $\lvert + \rangle$ and $\lvert - \rangle$ as the two
qubit ket states.
For a given pulse sequence with intervals $\tau_i$ between successive
pulses, the evolution operator is then
$\hat{U} = \sum_{\pm} \lvert \pm \rangle \hat{U}_{\pm} \langle \pm
\rvert$ (or $\hat{U} = \sum_{\pm} \lvert \mp \rangle \hat{U}_{\pm} \langle \pm
\rvert$ if there are an odd number of pulses)
with
\begin{equation}
\hat{U}_{\pm} = ... \exp{\left(-i \hat{\cal H}_{\mp} \tau_2\right)} 
\exp{\left(-i \hat{\cal H}_{\pm} \tau_1\right)}.
\label{evolOp}
\end{equation}
In order to characterize the coherence decay, we consider
the transverse component of the
qubit's expectation value; 
normalized to a maximum of one, the pulse sequence echo, $v_E$, is
defined in this way such that
$v_E = \left\| \left \langle \hat{U}_-^{\dag} \hat{U}_+ \right \rangle \right\|
=  \left\| \langle \hat{W} \rangle \right\|$
where $\hat{W} \equiv \hat{U}_-^{\dag} \hat{U}_+$, 
the $\langle ... \rangle$ denotes an appropriately weighted
average over the bath states (we use equal weights justified for
temperatures large compared to nuclear Zeeman energies), and $\|...\|$
is the magnitude of the resulting complex number.

{\it Uhrig series.}
The UDD sequence with $n$ pulses may be defined by~\cite{Uhrig}
\begin{equation}
\label{UhrigTau}
\tau_j = \frac{1}{2} \left[\cos{\left(\frac{\pi (j-1)}{n+1}\right)} -
  \cos{\left(\frac{\pi j}{n+1}\right)}  \right] t,
\end{equation} 
for $1 \leq j \leq n+1$ 
[corresponding to $\tau_j$ in Eq.~(\ref{evolOp})]
where $t$ is the total sequence time.
This series was shown~\cite{Uhrig} to optimally,
with respect to number of pulses, suppress decoherence
in the SB model.
We find, remarkably, that for any form of dephasing Hamiltonian, with
no assumptions about $\hat{\cal H}_{\pm}$, this sequence yields $v_E =
1 -\bigO{t^{2 n+2}}$.  Equivalently stated~\cite{MagnusFootnote}, $n$ pulses in
the UDD series removes the first $n$ orders of the Magnus
expansion of $v_E$!  In comparison with CDD, the number of required
pulses scales exponentially with respect to the orders of the Magnus
expansion that are cancelled~\cite{Khodjasteh}.
(It is important to note, however, that CDD can compete with UDD when
the Magnus expansion does not converge well).
We will return to a discussion of the universality
(Hamiltonian/model independence) of the UDD series at the end of this
Letter
after we demonstrate
its consequence for a model that is drastically different from SB.

{\it Spin boson versus spin bath.}
In the SB model, the spin qubit
interacts with a 
bath of non-interacting bosons:
${\cal H} = \sum_i \omega_i \hat{b}^{\dag}_i \hat{b}_i
+ \hat{S}_z \sum_i \lambda_i (\hat{b}^{\dag}_i + \hat{b}_i)$,
where $\hat{b}$ represents boson operators and $\hat{S}_z$ is the
$z$ spin operator for the central spin.
In contrast, the spin bath model treats interactions of a central
spin, such as a localized electron in a solid, with interacting
nuclear spins in a solid-state lattice.
Exemplifying the spin-bath model, SD is a dephasing of
the central spin as a result of fluctuations of the bath-induced
effective magnetic field caused by intra-bath interactions.
In the limit of a large applied field, the polarization of the central
electron spin and the nuclear spins must be individually preserved
(the electron having a gyromagnetic ratio that is typically $2000$
times larger than those of the nuclei), so that the Hamiltonian is
\begin{equation}
\label{H_SD}
\hat{\cal H} = \sum_n A_n \hat{S}_z \hat{I}_{nz} + 
\sum_{n \ne m} \left(b_{nm}
\hat{I}_{n-} \hat{I}_{m+} + c_{nm} \hat{I}_{nz} \hat{I}_{mz} \right).
\end{equation}



{\it Spectral diffusion: cluster expansion.} Despite the mesoscopic size of the solid-state baths that typically
contribute to SD, often involving many millions of
nuclear spins, it is feasible to compute
$v_E = \left\| \langle \hat{W} \rangle \right\|$ with $\hat{W} \equiv \hat{U}_-^{\dag} \hat{U}_+$ 
using a cluster expansion
that
breaks up the problem into manageable sub-problems that each involve
only a few nuclei.  (This expansion was 
successfully applied~\cite{witzelSD} 
to the problem of Si:P donor electron
SD yielding remarkable agreement~\cite{witzelAHF} with 
experiments~\cite{TyryshkinCoherence}.)
Consider expanding $\hat{W}$ such that
 $\hat{W} = \sum_{n=0}^{N} \clusterCountW{n}$ where $\clusterCountW{n}$ contains
contributions to $\hat{W}$ that involve $n$ separate clusters of
``operatively'' interacting nuclei.  
To be specific, the set of nuclei involved in a term of
$\clusterCountW{1}$ must all be connected together via factors of bilinear
interaction operators to form a single connected cluster.
Clusters have spatial proximity when interactions are local.
If it is possible to approximate $\langle \clusterCountW{1} \rangle$
by only including clusters up to some small size that is much less
than the number of nuclei in the bath, $N$, and if the initial bath
state is effectively uncorrelated (e.g., a random bath), then
$\langle \clusterCountW{n} \rangle \approx \langle \clusterCountW{1}
\rangle^n / n!$.
In this ``cluster approximation,''
$v_E = \left\| \langle \hat{W} \rangle \right\| \approx 
\exp{\left(\Real\left\{ \langle \clusterCountW{1} \rangle \right\}\right)}$.
This approximation is extremely useful because it is possible to treat
$\langle \clusterCountW{1} \rangle$ perturbatively in cases where the
perturbation would fail for $\langle \hat{W} \rangle$ directly due to
the vast numbers of multiple clusters involved in $\langle \hat{W}
\rangle$ (by definition $\clusterCountW{1} $ involves only single
clusters).
In our calculations, we approximate 
$\langle \clusterCountW{1} \rangle$ using a perturbation in cluster
size (we include contributions from successively increasing clusters
until convergence is achieved).  
The justification for this perturbation in cluster size is that each
additional nucleus in the cluster requires an additional bilinear
interaction factor.  An expansion in cluster size rather than a direct
expansion in orders of intra-bath interaction 
factors (e.g., diagramatically~\cite{SaikinLinkedCluster}) 
is simply more convenient.

{\it Spectral diffusion: time perturbation.}
In addition to the intra-bath perturbation, it is also possible for
$\langle \clusterCountW{1} \rangle$ to converge in a time expansion.
The time perturbation applies when the $\tau_j$ are 
small compared to all interactions time scales of the system.
The time perturbation is only really relevant if it is applicable on the
time scale of the decay (e.g., $T_2$ observed in spin echo).
Different clusters operate, in the sense of contributing to SD decay,
on very different time scales depending largely upon differences in
the HF interactions ($A_n$) among cluster nuclei, and these interactions are 
inhomogeneous over the bath.
When there are enough
clusters with enough influence operating on the shortest time scales
so that these clusters dominate the decay,
the decay time may be small compared to all interaction time scales
such that the time perturbation is relevant.
This, in
turn, depends on the distribution of the $A_n$ determined by the shape
of the electron wavefunction.
With these considerations and assuming that intra-bath interactions
are local (e.g., dipolar interactions), 
electrons in quantum dots 
with gaussian-shaped wavefunctions will tend to, in general, exhibit
short-time behavior SD on the decay time scale while donor-bound
electrons, with radially exponential wave functions,
will not~\cite{witzelWF}.
As an example of a situation in which the time perturbation is {\it
  not} appropriate,  the Hahn echo decay of donor-bound electrons in 
Si:P has the form
$\exp{(-\tau^{2.3})}$~\cite{witzelSD} which cannot be explained by any 
Magnus or time expansion.  The Hahn echo decay of GaAs quantum
dots, on the other hand, exhibit $\exp{(-\tau^{4})}$
behavior~\cite{witzelSD} explained by the lowest order of a time
perturbation expansion.

\begin{figure}
\includegraphics[width=3in]{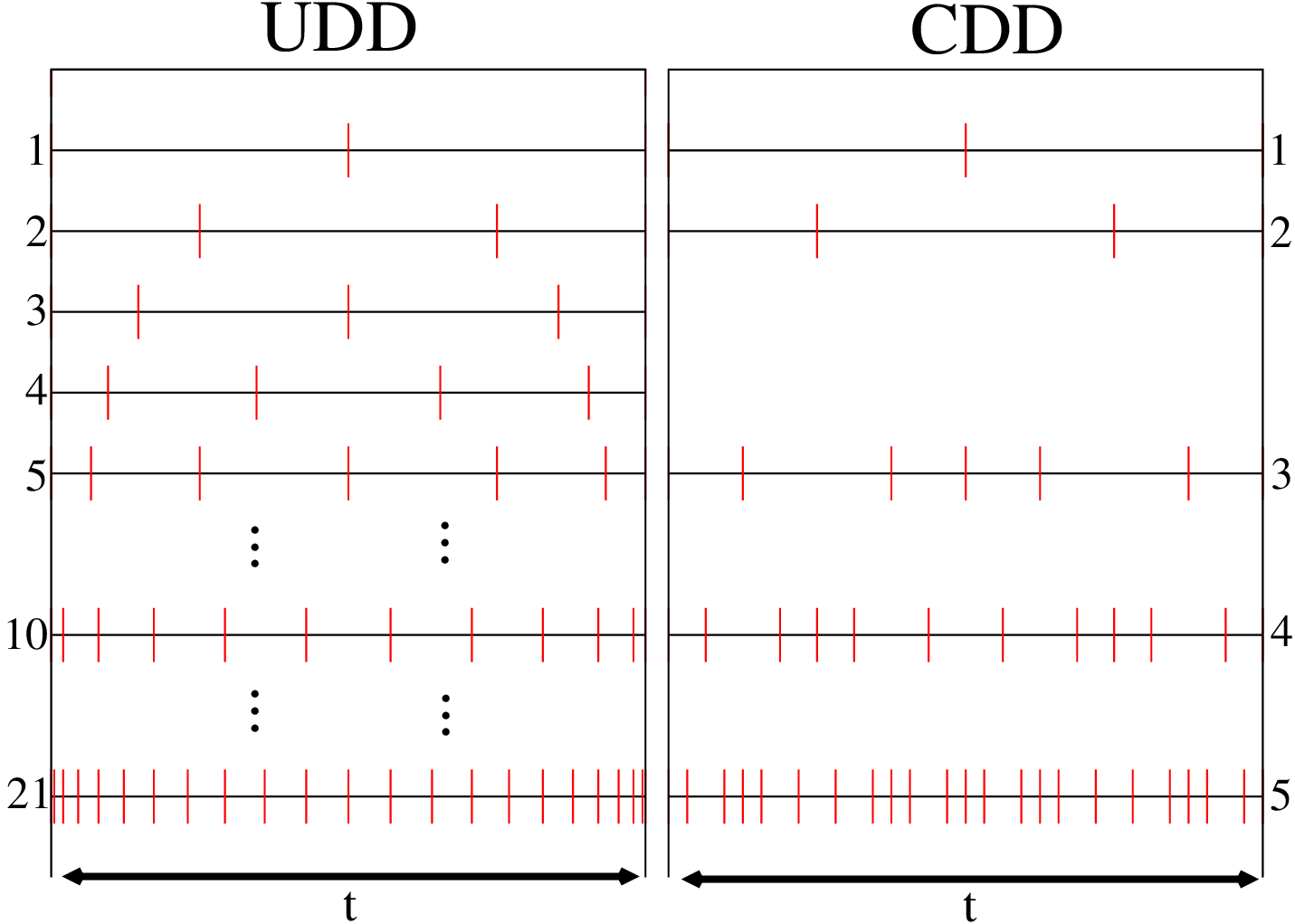}
\caption{
\label{figSequences}
(color online). 
Various UDD (left) and CDD (right) pulse
sequences in side-by-side comparison 
where red lines mark the time of each applied $\pi$ pulse.
The sequences are numbered by their order which corresponds,
respectively,
to the number of pulses (UDD) or the level of concatenation (CDD).
Orders of UDD and CDD with the same number of pulses are lined up horizontally.}
\end{figure}

{\it Spectral diffusion: pulse sequences.}
Using the cluster expansion technique, we are able to test the
effectiveness of DD strategies that use $\pi$-pulse
sequences in the real-world SD problem.  We
have previously theoretically verified~\cite{witzelCDD} the powerful effect of
SD suppression when applying
concatenations of the Hahn echo sequence
defined recursively by $\mbox{p}_l := 
\mbox{p}_{l-1} \rightarrow \pi \rightarrow \mbox{p}_{l-1} \rightarrow \pi$
with $\mbox{p}_0 := \tau$.  
These CDD sequences cancel successive
perturbative orders of both the intra-bath~\cite{witzelCDD} and
Magnus/time~\cite{Khodjasteh, yaoCDD, MagnusFootnote} 
expansions with each
concatenation.  
The main advantage of this sequence compared with UDD is that it
operates on the intra-bath perturbation which can be applicable on a
much longer time scale, for the time between pulses, than a time expansion 
(intra-bath coupling is on the order of $\mbox{ms}$
while HF interactions limits the time expansion on the order of 
$\mbox{$\mu$s}$).
Each concatenation, however, essentially doubles the
number of applied pulses, leading to exponential overhead.  
The main advantange of the UDD series [Eq.~(\ref{UhrigTau})]
is that it yields successive time expansion cancellations with each added pulse
for the SB model, a linear overhead!  
Figure~\ref{figSequences} shows a side-by-side comparison of UDD and
CDD sequences.


\begin{figure}
\includegraphics[width=3in]{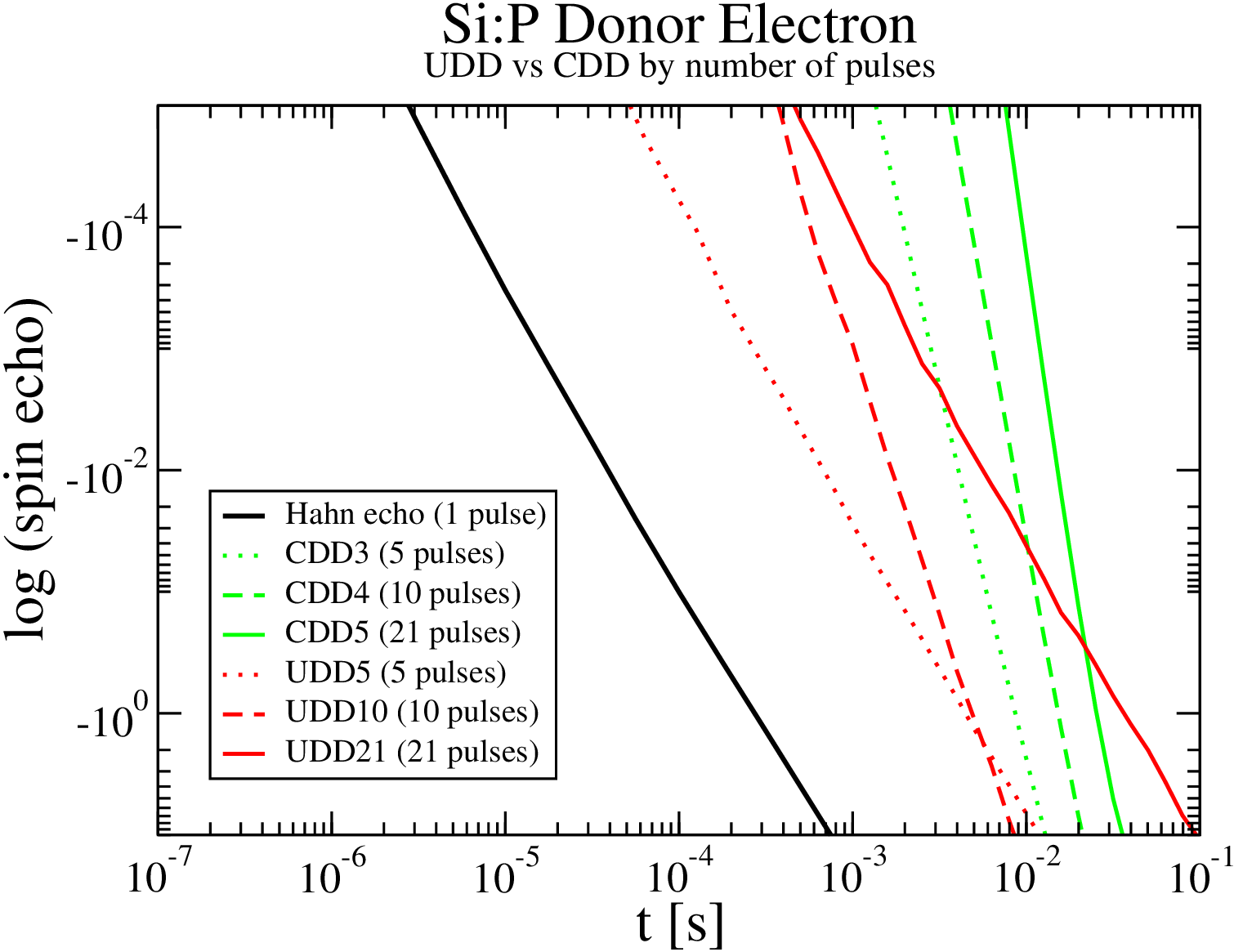}
\caption{
(color online). 
Numerical results showing the natural 
logarithm of echoes from CDD and UDD sequences
for Si:P donor electrons in natural Si.
Red (green) denotes UDD (CDD) sequence results; the Hahn echo
decay is shown as a solid black curve for reference.
Sequences with the same number of pulses have the same line pattern.
UDD has no particular advantage (in the high fidelity regime) to CDD
in Si:P,
but UDD with an even number of pulses performs better than those with
an odd number.
\label{figSi}}
\end{figure}

\begin{figure}
\includegraphics[width=3in]{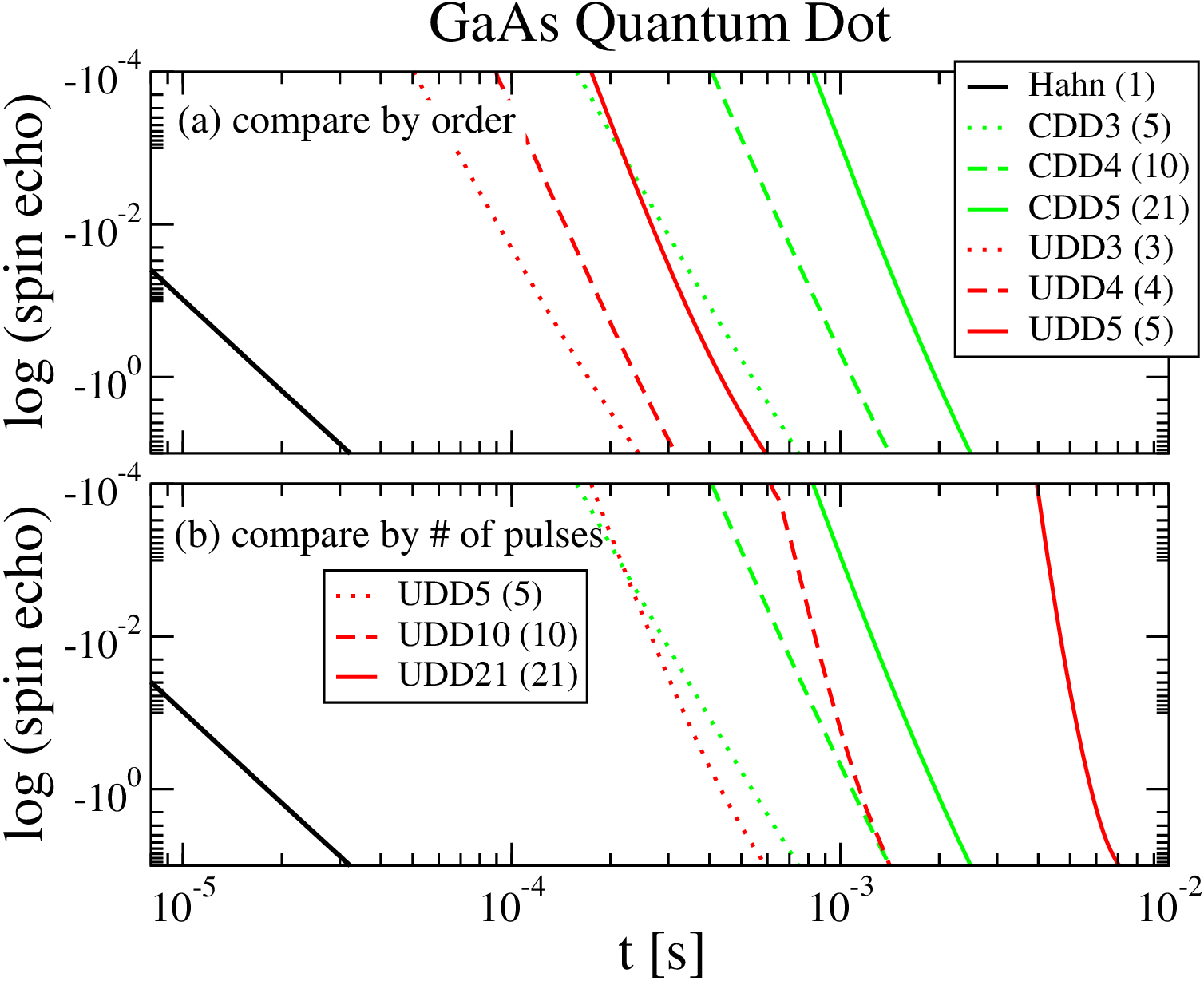}
\caption{
(color online). 
Numerical results showing the natural 
logarithm of echoes from CDD and UDD sequences
for quantum dots in GaAs with a $25~\mbox{nm}$
Fock-Darwin radius, $8.5~\mbox{nm}$ quantum well thickness.
Red (green) denotes UDD (CDD) sequence results; the Hahn echo decay
is shown as a solid black curve for reference.
(a) Sequences of the same order, having the same initial slope,
use the same line pattern. 
(b) Sequences with the same number of
pulses have the same line pattern.
Note that the Hahn (black) and CDD (green) curves in (a) and (b) are
the same, shown for comparison with UDD curves.
\label{figGaAs}}
\end{figure}

{\it Spectral diffusion: results.}
We show cluster expansion results (to the lowest non-trivial order in
the intra-bath perturbation), with coupling constants of
Eq.~(\ref{H_SD}) obtained using models described in Ref.~\cite{witzelSD}.
We compare the effects of CDD and UDD sequences on the coherence of 
both an electron bound to a P donor in 
natural Si [Fig. \ref{figSi}] and a quantum dot electron in GaAs
[Fig. \ref{figGaAs}], 
plotted as a function of the total time $t$ of one iteration
of the pulse sequence. 
In Si:P, where only the intra-bath
perturbation (and not the time perturbation) is applicable,
CDD maintains high fidelity (e.g.,
$10^{-4}$ decay) for a longer time than UDD.
The CDD sequences cancel out perturbative
orders of the intra-bath coupling  
with each concatenation~\cite{witzelCDD}; however, UDD sequences 
cancel only the first and, if there are an even number of pulses, 
second orders. 
In the GaAs quantum dot, in which time perturbation is valid, CDD still 
does better (maintains high fidelity longer)
than UDD for equal cancellations of order
[Fig.~\ref{figGaAs}~(a)]. However, 
for equal numbers of pulses, UDD preserves coherence far better than 
CDD [Fig.~\ref{figGaAs}~(b)].  
In this case, UDD 
cancels out perturbative orders of a time expansion (exhibited by the
successively increasing slopes of the curves) with each 
extra pulse, exactly what it was shown to do in the SB model. 
The only significant difference between Si:P and the GaAs quantum dot is the 
applicability of time perturbation expansion due to their respective
electron wavefunction shapes. 
Despite the stark, qualitative difference between SB and SD models,
UDD proves to be optimal, in the short time limit (where accessible), for both.

{\it Universality.}
What is most striking, however, is that
UDD is generically optimal for
 cancelling orders of a time perturbation
for {\it any} dephasing decoherence!
With full generality, let $\hat{\cal H}_{\pm} = \hat{X}_0 \pm
 \hat{X}_1$ without making any assumption about the
 commutation properties of $\hat{X}_0$ and $\hat{X}_1$.
This Hamiltonian, along with the pulse time intervals, $\tau_j$, for a
 given pulse sequence will determine the $\hat{U}_{\pm}$ operators
 that go into $v_E = \left\| \langle \hat{W} \rangle \right\|$ with
$\hat{W} \equiv \left[\hat{U}^-\right]^{\dag} \hat{U}^+$.  
Taking the real part of $\hat{W}$ yields a more convenient expression,
 $\Real{\left\{ \langle \hat{W} \rangle \right\}} = 1 - \langle
 \Delta^{\dag} \Delta \rangle / 2$ with $\Delta = \hat{U}^{+} -
 \hat{U}^{-}$.
Since $v_E > \Real{\left\{ \langle \hat{W} \rangle \right\}}$, it only
 serves to make a more pessimistic estimate to ignore the imaginary
 part of $\langle \hat{W} \rangle$; we may therefore restrict our attention to the real
 part in our perturbative analysis.

Expanding the exponentials of $\hat{U}_{\pm}$ [Eq.~(\ref{evolOp})] and
collecting terms with common sequences of $\hat{X}_{0,1}$ operators,
we define $C_{i_1, i_2, ...}$ as the coefficient of a term in $\hat{\Delta}
= \hat{U}_+ - \hat{U}_-$ for a corresponding
sequence of operators such that
\begin{equation}
\hat{\Delta} = \sum_{m=0}^{\infty} \sum_{i_1, i_2, ..., i_m = 0}^{1}
(-i)^m C_{i_1, i_2, ..., i_m} \hat{X}_{i_1} \hat{X}_{i_2}
... \hat{X}_{i_m}.
\end{equation}
These $C_{...}$ coefficients may be written as a sum of products 
of the $\tau_j$ pulse delay times, 
defined by Eq.~(\ref{UhrigTau}) for a sequence
of $n$ pulses, by expanding Eq.~(\ref{evolOp}).
It so happens that for any $m \leq n$, all $C_{i_1, i_2, ..., i_m}
\sim \bigO{t^m}$
coefficients are identically zero!  We have explicitly
proven this for (up to) $n \leq 9$ 
using computer integer arithmetic by representing each
$C_{...}$ as a polynomial of $\alpha = \exp{\left(i \pi /
  (n+1)\right)}$, exploiting the fact that $\alpha^{n+1} = \alpha^{-n-1} =
-1$, and noting that two or more points placed symmetrically 
(i.e., equally spaced)
around a circle in the
complex plane sum identically to zero.
We conjecture that
this result is generically true for all integer $n$, and 
numerical calculations (up to $n=14$) are consistent with this assertion.
This finding is remarkable!  The UDD series satisfy 
an exponentially growing set of $C_{...} = 0$ non-linear equations
with degrees of freedom ($\tau_j$) that merely grow linearly (one for
each pulse).

{\it Conclusion.}
We conclude by emphasizing our key finding that
the UDD~\cite{Uhrig} sequences
restore coherence optimally and generically in a model independent
manner through
successive cancellations in orders of a
time/Magnus\cite{MagnusFootnote}
 perturbation expansion with a low 
overhead of a single pulse for each order of cancellation.
While we assume ideal, instantaneous pulses in our analysis,
careful pulse shaping can theoretically\cite{UhrigPulseShaping} 
mitigitate effects of the finite width in experimentally applied pulses.
The universal Hamiltonian-independent applicability of this series of
pulse sequences, originally proposed within the narrow constraint of a
spin boson model, is simply miraculous.

We thank Professor G\"{o}tz Uhrig for bringing his optimal sequences
with linear scaling (in the number of pulses) to our attention,
and {\L}ukasz Cywi{\'n}ski for extremely helpful advice in the
presentation of this paper.
We thank Professor Deepak Dhar for bringing Ref.~\cite{Dhar} to our attention.
This work is supported by LPS-NSA and ARO-DTO.


\begin{thebibliography}{99}
\bibitem{MagnusFootnote}Note that there is a direct correlation
  between cancelations in a Magnus\cite{Magnus} or time expansion
  arising from the simple fact that $\exp{(t^n)} = 1 + O(t^n)$.
\bibitem{footnote}It has recently come to our attention that the
  universality of the first five UDD sequences (i.e., up to five
  pulses) was known~\cite{Dhar} before the full UDD series~\cite{Uhrig} was discovered.
\end{thebibliography}
\end{document}